# The Search for Gravitational Waves

**Jim Hough and Sheila Rowan**
Department of Physics and Astronomy, University of Glasgow, Glasgow, G12 8QQ, UK

**B.S. Sathyaprakash**
  School of Physics and Astronomy, Cardiff University, 5 The Parade, Cardiff, CF24 3YB, UK

**Abstract**
Experiments aimed at searching for gravitational waves from astrophysical sources have been under development for the last 40 years, but only now are sensitivities reaching the level where there is a real possibility of detections being made within the next five years. In this article a history of detector development will be followed by a description of current detectors such as LIGO, VIRGO, GEO 600, TAMA 300, Nautilus and Auriga. Preliminary results from these detectors will be discussed and related to predicted detection rates for some types of sources. Experimental challenges for detector design are introduced and discussed in the context of detector developments for the future.

## 1. Introduction

For many years physicists have been facing up to the exacting experimental challenge of searching for gravitational waves. Predicted by General Relativity to be produced by the acceleration of mass [1], but considered by early relativists to be transformable away at the speed of thought they have remained an enigma ever since.

What are gravitational waves? This will be discussed more fully in the next section but they can be thought of as ripples in the curvature of space-time or as tiny fluctuations of the direction of *g*, the acceleration due to gravity, on Earth.

Why are we interested in their detection? To some extent to verify the predictions of General Relativity – although given the success of the other predictions of General Relativity being verified, it will be a major upset if gravitational waves do not exist! More importantly we want to use gravitational waves as a tool for looking into the heart of some of the most violent events in the Universe and so start a new branch of astronomy.

However to reach this point very sensitive detectors have to be developed and it is interesting to see the extent to which Einstein's other work - on Brownian motion [2] and the photoelectric effect [3] for example - is of relevance to the experimental field.

## 2. Gravitational Waves

To gain an impression of the nature of gravitational waves it is easiest to use the description of space-time from Special Relativity [4] where the proper distance between two neighbouring points in flat space time is given as $ds^2 = -c^2 dt^2 + dx^2 + dy^2 + dz^2 = \eta_{\mu\nu} dx^\mu dx^\nu$ where $\eta_{\mu\nu}$ is known as the Minkowski metric tensor and *c* is the speed of light.

General Relativity predicts that space-time is curved by the presence of mass and more generally $ds^2 = g_{\mu\nu} dx^\mu dx^\nu$ where the detail of the space-time curvature is contained in the metric tensor $g_{\mu\nu}$.

It is simplest to consider the case where the gravitational fields are very weak, and so the curvature of space is small, then $g_{\mu\nu} = \eta_{\mu\nu} + h_{\mu\nu}$ where $h_{\mu\nu}$ represents a small perturbation of the metric away from that for flat space-time.

Postulating that this perturbation might be sinusoidal in nature and working in a coordinate system defined by the trajectories of freely-falling test masses, Einstein's field equation yields a wave equation of the form $\left(\nabla^2 - 1/c^2 \frac{d^2}{dt^2}\right) h_{\mu\nu} = 0$ where the amplitude of the wave is related to the perturbation of the metric which is really the amplitude of the curvature of space-time.

Gravitational effects are tidal by nature and the gravitational wave amplitude, *h*, can usually be interpreted as a physical strain in space or, more precisely, $h = 2\delta l / l$ where $\delta l$ is the change in separation of two masses a distance *l* apart.

### 2.1. Generation and detection of gravitational waves

Gravitational waves are produced when mass undergoes acceleration, and thus are analogous to the electromagnetic waves that are produced when electric charge is accelerated. However the existence of only one sign of mass, together with law of conservation of linear momentum, implies that there is no monopole or dipole gravitational radiation. Quadrupole radiation is possible and the magnitude of *h* produced at a distance *r* from a source is proportional to the second time derivative of the quadrupole moment of the source and inversely proportional to *r*, while the luminosity of the source is proportional to the 'square' of the third time derivative of the quadrupole moment.

For quadrupole radiation there are two 'orthogonal' polarisations of the wave at 45 degrees to each other, of amplitude $h_+$ and $h_x$, and each of these is equal in magnitude to twice the strain in space in the relevant direction.

The effect of the two polarisations on a ring of particles is shown in figure 1, and from this the principle of most gravitational wave detectors – looking for changes in the length of mechanical systems such as bars of aluminium or the arms of Michelson type interferometers – can be clearly seen.

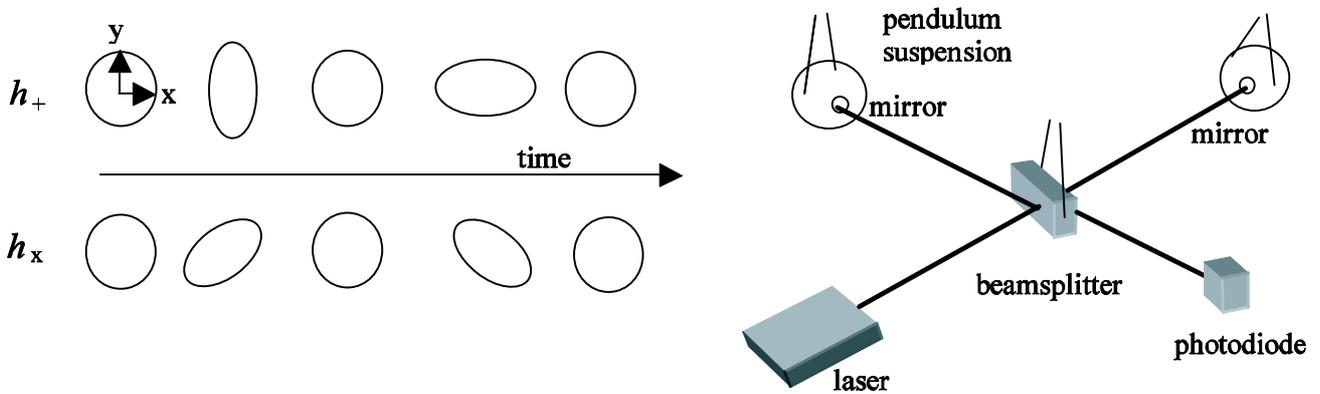

**Figure 1.** Schematic diagram of how gravitational waves interact with a ring of matter. The 'quadrupole' nature of the interaction can be clearly seen, and if the mirrors of the Michelson Interferometer on the right lie on the ring with the beamsplitter in the middle, the relative lengths of the two arms will change and thus there will be a changing interference pattern at the output.

The problem for the experimental physicist is that the predicted amplitudes or strains in space in the vicinity of the Earth caused by gravitational waves from astrophysical events are extremely small, of the order of $10^{-21}$ or lower [5]. Indeed current theoretical models of the event rate and strength of such events suggest that in order to detect a few events per year - from coalescing neutron star binary systems for example - an amplitude sensitivity close to $10^{-22}$ over approximately 1000 Hz is required, and thus detector noise levels must have an amplitude spectral density lower than $\sim 10^{-23}/\sqrt{Hz}$. Signal strengths at the Earth, integrated over appropriate time intervals, for a number of sources are shown in figure 2.

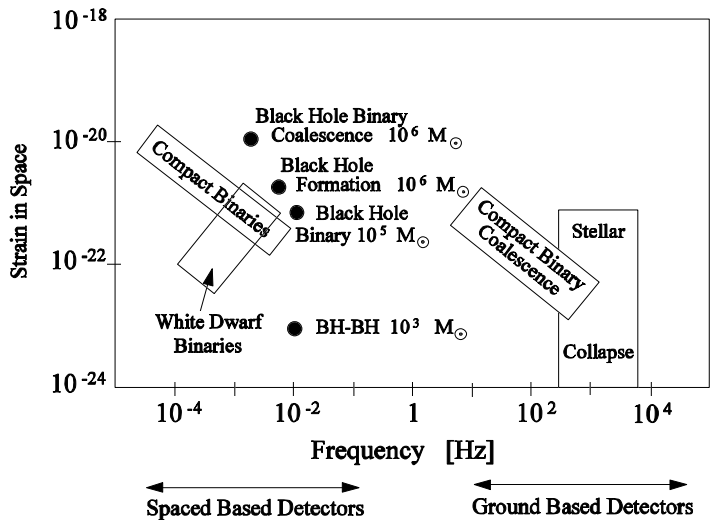

**Figure 2.** Predicted signal strengths for a number of possible sources of gravitational waves

The weakness of the signal means that noise sources like the thermal motion of molecules in the detector (thermal noise), seismic or other mechanical disturbances, and noise associated with the detector readout, whether electronic or optical, must be reduced to a very low level. For signals above ~10 Hz, ground based experiments are possible, but for lower frequencies where local fluctuating gravitational gradients and seismic noise on Earth become a problem, it is best to develop detectors for operation in space [6, 7].

*2.2 Sources of gravitational waves*

Gravitational wave detectors will uncover dark secrets of the Universe by helping us to study sources in extreme physical conditions: strong non-linear gravity and relativistic motion, extremely high density, temperature and magnetic fields, to list a few.

Gravitational wave signals are expected over a wide range of frequencies, from $10^{-17}$ Hz in the case of ripples in the cosmological background to $10^3$ Hz when neutron stars are born in supernova explosions. Because of the very weak nature of gravity and lack of dipole radiation, the efficiency of converting mechanical energy in a system into gravitational radiation is very low and thus signals produced by accelerating systems tend to be very weak. Indeed the only sources of gravitational waves (GWs) that are likely to be detected are astrophysical, where there are potentially huge masses accelerating very strongly. There are many sources of great astrophysical interest including black hole interactions and coalescences, neutron star coalescences, low-mass X-ray binaries such as Sco-X1, stellar collapses to neutron stars and black holes (supernova explosions), rotating asymmetric neutron stars such as pulsars, and processes in the early Universe. The binary coalescences and likely detection rates will be discussed in detail in section 6.

**3. History**

There appears to have been little interest in the experimental detection of gravitational radiation for forty-five years after their prediction. However in the late 1950s this changed with Joseph Weber of the University of Maryland suggesting the design of some relatively simple apparatus for their detection [8, 9]. This apparatus in its later stages consisted of an aluminium bar of mass approximately one ton with piezoelectric transducers bonded around its centre line. The bar was suspended from anti-vibration mountings in a vacuum tank. By means of the amplified electrical signals from the transducers Weber monitored the amplitude of oscillation of the fundamental mode of the bar. A gravitational wave signal of suitable strength would be expected to change the amplitude or phase of the oscillations in the bar. In the 1969/70 period Weber operated two such systems one at the University of Maryland and one at the Argonne National Laboratory and observed coincident excitations of the bars at a rate of one event per day [10, 11]. These events he claimed to be gravitational wave signals.

However other experiments – at Moscow State University [12], Yorktown heights [13], Rochester [14], Bell Labs [15], Munich [16] and Glasgow [17] - failed to confirm Weber's detections. The detector at the Max-Planck-Institut fuer Physik und Astrophysik in Munich is shown in Fig 3(a). Several years of lively debate about the interpretation of Weber's results followed, the outcome being a somewhat predictable standoff between Weber and the rest of the community. An analysis of detector sensitivity of the Weber bar design suggested that the sensitivity was approximately $10^{-16}$ for millisecond pulses. However an event rate of one per day resulting from events at the centre of the galaxy - as claimed by Weber - corresponded to a very high loss of energy, and thus mass, from the galaxy, so high in fact that changes in the position of the outermost stars should have been visible due to a reduction in gravitational force towards the galactic centre [18]. A solution suggested for this – beaming of the energy in a narrow cone so that each detected event implied much less overall energy loss - was discussed by many authors but did not receive wide acceptance.

Thus the field had to find a new way forward, the driving force being the need to improve detector sensitivity. There are two fundamental ways to improve sensitivity in any detector system. The first is to reduce the background noise level, and the second is to increase the signal size. Post-Weber detector developments followed both of these routes. In the case of resonant bar detectors the main limitations to sensitivity were from the thermal excitation of its normal modes and from electronic noise in the detection system [19] and thus the most obvious way to reduce detector noise level was to reduce the temperature of the aluminium bars systems.

*3.1. Low temperature resonant bars*

Reduction of the operating temperature was the direction followed by Weber himself. He was joined in this by research groups at Stanford University [20, 21], Louisiana State University (Allegro detector) [21, 22], the University of Rome (Explorer and Nautilus detectors) [23,24], the University of Western Australia (Niobe) [25], and more recently a collaboration of the Universities of Trento and Padua (Auriga) [26]. Bars were of the order of a ton in weight and with the exception of the UWA bar were aluminium. UWA used niobium, a material that, in principle, allowed better thermal noise performance to be achieved. The Auriga detector at Legnaro (Trento/Padua) is shown in figure 3(b).

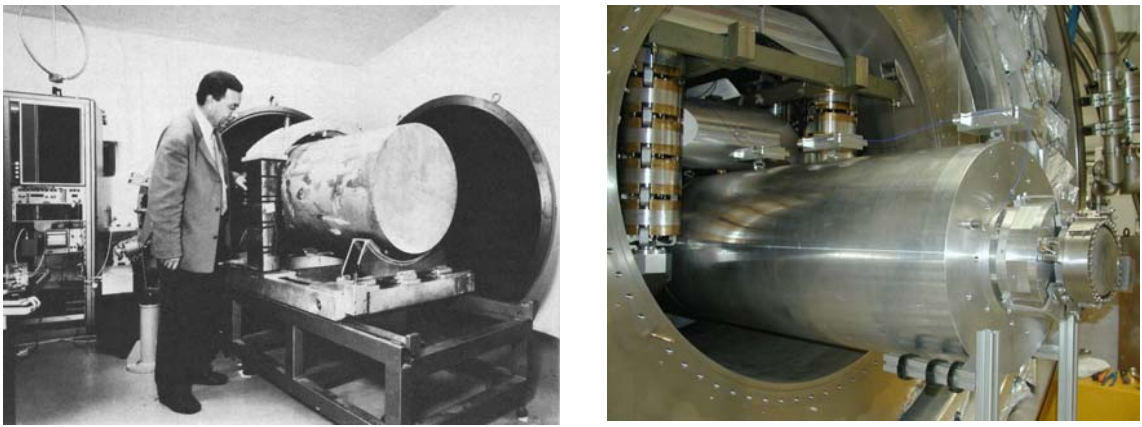

**Figure 3.** (a) Room temperature bar detector in Munich (with H. Billing) and (b) Low temperature bar detector, Auriga, at Legnaro.

A number of experiments have been carried out over the last twenty-five years with these bar detectors, there being long periods where they were working together - sometimes in pairs, sometimes with more detectors - and many papers have been published, [21-27] for example. However despite occasional reports that some coincident events had been observed by a number of groups no definitive evidence for the existence of gravitational waves has yet been put forward. In 2003 analysis of data over a ninety-day period from 2001 from the Nautilus detector in Rome and the Explorer detector in CERN yielded eight coincident events each occurring between hours 3 and 5 in the sidereal day [28]. Analysis of the time of reception of the events together with an analysis of the reception sensitivity pattern of the detectors suggested that if they were gravitational wave events they were originating from the plane of the galactic disk. There has been significant discussion of the statistical significance of the events by Sam Finn of Penn State and colleagues [29] and a lively response from the authors of the relevant paper [30]. The Rome group do not claim that they have detected gravitational waves rather that they have observed some interesting events. They have been looking closely at the energetics of the signals. Eugenio Coccia of the Rome group concludes that if the signals are from gravitational waves, standard sources such as coalescing compact binaries in the galaxy would not happen often enough; rather some more exotic source such as one or more repeating bursters in the plane of the galaxy would be needed [31]. Analysis of a further year of data is currently underway and may shed further light on the situation although preliminary indications from the Rome group suggest that the outcome is less than clear. Indeed it may be that we need results from the large interferometric detectors discussed below to help us understand the significance of these results.

The current sensitivity of low temperature bars is 1000 times better in terms of gravitational wave amplitude than the initial Weber type detectors. Further improvements are promised with operation at millikelvin temperatures and with better low temperature amplifiers becoming available for the sensing systems. Further, new detector designs are evolving in which spheres of low temperature material replace the cylindrical bars. The spheres are instrumented in such a way that the effect of the gravitational wave on a number of modes of the sphere can be detected leading to higher effective cross section and directional sensitivity. Prototype low temperature spherical systems are being developed in the Netherlands (MiniGRAIL [32]) and in Brazil (Mario Schenberg [33]).

Until recently the bar detectors have had a very narrow operational bandwidth, ~ 1 Hz, but the present generation of bars as exemplified by Auriga has a bandwidth of approximately 80 Hz with sensitivity approaching $10^{-21}/\sqrt{Hz}$. This bandwidth will be further increased if a new nested detector design as proposed by Massimo Cerdonio and colleagues of the Universities of Padua and Trento is built [34]. In this nested design one mass has a lower resonant frequency than the other, and thus between the resonances, the masses respond in antiphase to a gravitational wave signal hence enhancing the effect. Developing sensors to detect the relative motion of the components of the nested system is an interesting challenge for the future.

*3.2. Long baseline interferometric detectors*

The other approach to enhancing detector sensitivity, adopted by the German and UK groups, was to aim for signal enhancement by moving the test masses apart and using laser interferometry to sense the relative motion. This idea was not new but had been awaiting the availability of high enough powered lasers to have the potential of high sensitivity.

Weber's early work on methods of detecting gravitational waves encouraged others to devise potentially sensitive detector designs. Among these were Gertsenshtein and Pustovoit from Moscow who proposed in 1962 that one could look for small shifts in the fringe pattern in a Michelson interferometer formed between freely hung mirrors [35]. Such an interferometric arrangement is shown in Figure 1 with a laser replacing the white light source originally suggested. An interferometer of this type - with rubber isolating stacks replacing the pendulums - was constructed by Robert Forward at the Hughes aircraft corporation in the late 1960s following suggestions to do this from his former Ph.D. supervisor Joseph Weber. The arm length of the interferometer was 2 m and it was illuminated with light power of a few mW from a helium-neon laser. To a large extent Forward's interferometer was limited in performance by the statistics of the photoelectrons released in the photodetector by the light from the fringes, with a strain sensitivity of ~ 6 x $10^{-15}/\sqrt{Hz}$ being achieved at frequencies near a kHz. Integrated over a bandwidth of a kHz this would allow pulses of duration 1 msec to be detected at a level greater than 2 x $10^{-13}$ [36]. As discussed earlier the effect of a gravitational wave is likely to be at least eight to nine orders of magnitude smaller than this and thus much development of these interferometric detectors was required. Forward, with colleagues, did improve his detector by nearly two orders of magnitude by folding the optical path and increasing the laser power [37]. Folding of the optical path by means of a Herriott delay line - where it is possible to have many hundred foldings - was suggested by Rai Weiss at MIT [38] and incorporated into a short interferometric prototype detector constructed at MIT thereafter.

By this time laser strainmeters of up to 30 m arm length had been demonstrated by Levine and Hall in the Poorman mine in Colorado [39] and had used to carry out a search for periodic gravitational waves from the Crab pulsar through their effect on the Earth [40]. Further, much higher power argon-ion lasers were becoming available and separated mass detectors using laser interferometry between masses separated by tens of meters began to look to be a realistic way ahead.

In Germany a 3 m prototype using optical delay lines was followed by a 30 m instrument developed during the early 1980s at the Max Planck Institute for Astrophysics in Garching [41, 42]. In the UK a 1 m instrument (1977) using a different form of multi-beam optical system was followed during the early 1980s by a 10 m interferometer using Fabry-Perot cavities in the arms [43, 44, 45]. Then a 40 m instrument was developed in Caltech as a spin-off from the Glasgow interferometer when Ronald Drever moved to Caltech [46, 47]. Two prototypes, one using delay lines and the other Fabry-Perot cavities, were also built in Japan [48, 49]. All of these instruments used multiwatt argon-ion lasers and the majority achieved displacement sensitivities of better than $10^{-18}$ m/$\sqrt{Hz}$ over a frequency range of a few hundred Hz to a kHz.

At this stage in many countries the technology was considered sufficiently mature for a strong case to be made for the construction of detectors of much longer baseline, detectors that should be capable of having a real possibility of detecting gravitational waves.

Thus an international network of gravitational wave detectors came into being. The American LIGO project, which sprang from the MIT and Caltech prototypes, comprises two detector systems with arms of 4 km length, one in Hanford, Washington State, and one in Livingston, Louisiana [50]. One half-length, 2 km, interferometer has also been built inside the same evacuated enclosure at Hanford. A birds-eye view of the Hanford site showing the central building and the directions of the two arms is shown in figure 4. Construction of LIGO began in 1996 and progress has been outstanding [51] with one of the LIGO detectors

– the Hanford 4 km instrument – being within a factor of two of design sensitivity over much of its frequency range at the present time (Summer 2004) [52].

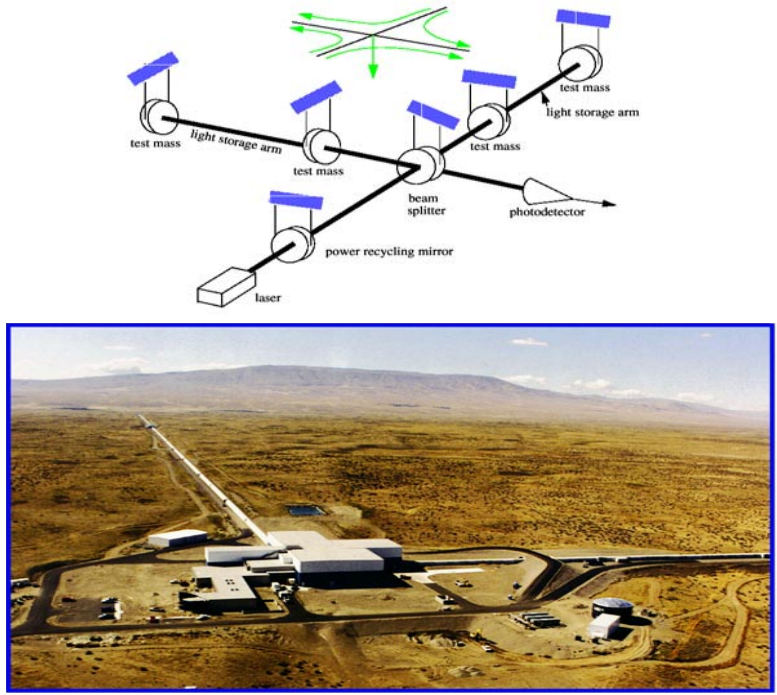

**Figure 4.** Schematic diagram and birds eye view of LIGO (Hanford).
Image courtesy of the LIGO Science Collaboration

The French/Italian VIRGO detector [53] of 3 km arm length at Cascina near Pisa is designed to have exceptionally good low frequency performance, down to 10 Hz, and is close to completion. The Japanese TAMA 300 detector, which has arms of length 300 m, is operating at the Tokyo Astronomical Observatory [54].

All the systems mentioned above are designed to use resonant cavities in the arms of the detectors and also place a mirror of finite transmission between the laser and the interferometer to impedance match one to the other and hence enhance the laser power at the beamsplitter of the interferometer. This technique is known as power recycling [55, 56]. They also use standard wire sling techniques for suspending the test masses. However the German/British detector, GEO 600, is somewhat different [57]. It makes use of a four-pass delay-line system with advanced optical signal enhancement techniques where a mirror at the output recycles the signal sidebands back into the interferometer. (This technique is known as signal recycling [58, 59, 60].) GEO utilises very low loss fused silica suspensions for the test masses (see section 4.2), and is expected to reach a sensitivity at frequencies above a few hundred Hz close to those of VIRGO and LIGO when they are in initial operation. GEO is now fully built and its sensitivity is being continuously improved. Currently (October 2004) it is within a factor of ten of design sensitivity over much of its frequency range.

Three science runs of increasing sensitivity, ranging from 17 to 70 days in length, have so far been carried out with these new interferometric detectors. All have involved the LIGO detectors, and two have involved the GEO and TAMA detectors. The Allegro bar detector in Louisiana has also taken part in the latest of these runs. From the first science run, upper limit results on the signals from a number of potential sources such as pulsars, coalescing compact binary stars, as well as on burst events and the level of a stochastic background, have been set [61, 62, 63, 64]. Results from the second run are about to be published, and those from the third run are being analysed.

During the next few years we can expect to see a series of increasingly sensitive searches for gravitational wave signals at a sensitivity level of approximately $10^{-21}$ for millisecond pulses or close to $10^{-26}$ for pulsars, to take two examples. This latter level equivalent to the neutron star having an ellipticity of $\sim 10^{-8}$ is at an astrophysically feasible level and thus the detection of gravitational waves from pulsars in the near term is a real possibility. Further, the recent discovery of another compact binary system in the galaxy - the double pulsar J0737-3039 – has improved the statistics for the expected rate of binary coalescences by a significant factor, implying that the most probable rate of binary neutron star coalescences detectable by the LIGO system now lies between one per 10 years and 1 per six hundred years [65]. Many people expect the rate of binary black hole coalescences to be even higher. Binary coalescences are discussed in more detail in section 6.

Detection at the level of sensitivity of the initial detectors is no way guaranteed; thus improvement of the order of a factor of 10 in sensitivity of the current interferometric detectors is essential to allow compact binary coalescences at least to be seen at a detectable level. Indeed, plans for an upgraded LIGO, Advanced LIGO, are already mature and the project has recently been approved by the National Science Board in the USA (October 2004). Plans are also well advanced for an underground detector with cooled test masses to be built in Japan [66]. The baseline design for advanced LIGO incorporates 40 kg sapphire test masses, suspended by fused silica fibers or ribbons, along with an improved seismic isolation system, increased laser power, of the order of close to 200 W, and signal recycling [67]. The upgrade is now expected to commence in 2009 and it is exciting to note that the most probable rate of detectable binary neutron star coalescences is now expected to be in the range of 10 to 500 per year [65]. The noise anatomy of Advanced LIGO is shown in figure 5.

**4. Limiting noise sources: the experimenter's challenge**

In this section we discuss the main noise sources, which limit the sensitivity of ground-based interferometric gravitational wave detectors. Fundamentally it should be possible to build systems using laser interferometry to monitor strains in space which reach or even supercede the Standard Quantum Limit (SQL) i.e. the limit set by the Heisenberg Uncertainty Principle. Indeed the proposed performance for Advanced LIGO is close to this limit at mid-frequencies. The SQL and related issues will be discussed in a later section

However there are other practical issues that must be considered. Fluctuating gravitational gradients pose one limitation to the interferometer sensitivity achievable at low frequencies, and it is the level of noise from this source which dictates that experiments to look for gravitational wave signals below 10 Hz or so have to be carried out in space [68, 69, 70, 71]. While there are schemes to monitor such gradients and cancel out their effects on the interferometers [72] we will not discuss them further here.

In general, for the practical building of ground based detectors the most important limitations to sensitivity result from the effects of seismic and other ground-borne mechanical noise, thermal noise associated with the test masses and their suspensions, shot noise in the photocurrent from the photodiode which detects the interference pattern, and radiation pressure recoil effects on the interferometer mirrors, these last two being intimately related with quantum limits to performance. The significance of each of these sources for present and future interferometric detectors will be briefly reviewed.

*4.1 Seismic noise*

Seismic noise at a reasonably quiet site on the Earth follows a spectrum in all three dimensions close to $10^{-7} \times f^{-2}$ m/$\sqrt{\text{Hz}}$ (where here and elsewhere we measure $f$ in Hz). If the disturbance to each test mass must be less than $3 \times 10^{-20}$ m/$\sqrt{\text{Hz}}$ at, for example, 30 Hz then the reduction of seismic noise required at that frequency in the horizontal direction (along which the gravitational wave induced strains are to be sensed) is thus greater than $10^9$. Since coupling of vertical noise through to the horizontal axis is typically of the order of $10^{-3}$, a significant level of isolation has to be provided in the vertical direction also. Isolation in the horizontal direction can be provided in a relatively simple way by making use of the fact that, for a simple pendulum system, the transfer function from the suspension point to the pendulum mass falls off as $f^{-2}$ above the pendulum resonance. In a similar way vertical isolation can be achieved by suspending a mass on a spring. In the case of the VIRGO detector system the design allows operation to below 10 Hz by adopting a seven stage horizontal pendulum arrangement with six of the upper stages being suspended by cantilever springs to provide vertical isolation [73]. Similar systems have been developed by DeSalvo at Caltech [74] now for installation in the Japanese TAMA detector.

For the GEO 600 detector, where operation down to 50 Hz is sought, a triple pendulum system is used with the first two stages being hung from cantilever springs to provide the vertical isolation necessary to achieve the desired performance. This arrangement is then hung from a plate mounted on passive 'rubber' isolation mounts and on an active (electro-mechanical) anti-vibration system [75, 76]. An extension of this system is being developed for Advanced LIGO. In this case a quadruple pendulum system [77] is being utilized.

In order to reduce unwanted motions at the pendulum frequencies, active damping of the pendulum modes has to be incorporated, and to reduce excess motions at low frequencies around the micro-seismic peak, low frequency isolators have to be incorporated. These low frequency isolators can take different forms. In the case of the VIRGO system, tall inverted pendulums are used with seismometer/actuator feedback in the horizontal direction and cantilever springs whose stiffness is reduced by means of attractive forces between magnets for the vertical direction [78]. Roberts mechanical linkages in the horizontal and torsion bar/Euler spring arrangements in the vertical are adopted in an Australian design [79] for the high power test facility operated by ACIGA at Gingin [80], and seismometer/actuator systems below the passive metal/rubber stages in GEO 600 are used [81, 82]. For Advanced LIGO a new sophisticated double stage, six degree of freedom active isolation system is under development [83].

*4.2 Thermal noise*

Thermal noise associated with the mirror masses and the last stage of their suspensions is, and is likely to continue to be, one of the most significant noise source at the low frequency end of the operating range of long baseline gravitational wave detectors [84].

Current suspension designs have been based predominantly on modeling the behaviour of the resonant modes of test masses and suspensions as damped harmonic oscillators. The power spectral density of thermal displacement noise, $S_x(f)$, associated with a mode of resonant frequency $f_0$ can then simply be written as [84]:

$$S_x(f) = \frac{k_B T f_0^2 \phi(f)}{2\pi^3 fm\left[\left(f_0^2 - f^2\right)^2 + f_0^4 \phi^2(f)\right]}$$

where $\phi(f)$ is the mechanical dissipation or loss factor of the oscillator of mass $m$, at temperature $T$. Inspection of the equation above indicates that at the resonant frequency $f_0$ the thermal displacement is very large for $\phi(f_0) \ll 1$ or equivalently for $Q \gg 1$ where $Q = 1/\phi(f_0)$. However at frequencies far from resonance the thermal displacement is proportional to $\sqrt{\phi(f)}$. The test masses and their suspensions in interferometric detections have thus been fabricated from materials of low mechanical dissipation and designed to have, where possible, resonant modes outwith the frequency band of interest for gravitational wave detection. In particular the longitudinal pendulum mode of a suspension is typically ~1 Hz and the internal resonant modes of the test mass mirrors are ~10's of kHz. However the transverse 'violin' modes of the fibres suspending the test masses are typically ~100's of Hz and thus appear in the detection band.

The off-resonance thermal noise of the pendulum and violin modes results from dissipation associated with the flexing of the suspension wires or fibres. High strength carbon steel wires are used in the LIGO, VIRGO and TAMA suspensions. In contrast, as mentioned earlier, the GEO 600 test mass suspensions use cylindrical fused silica suspension fibres in the final stage since the intrinsic dissipation of silica is substantially lower than that of steel, [85, 86, 87, 88, 89] whilst having comparable tensile strength [90].

It is important to note that for both the pendulum and violin modes, the resulting thermal noise is reduced over that expected simply from consideration of the internal dissipation of the fibre material through the fact that part of the potential energy associated with the flexing of each fibre is stored as gravitational potential energy in the Earth's loss-less gravitational field, thus 'diluting' the suspension fibre dissipation [84]. Through careful suspension design the dissipation factors of the in-band violin modes can thus be made low enough that the associated thermal displacement occupies very little of the detector frequency band.

Over much of the detector frequency band the off-resonance thermal noise from the test mass mirrors has a more significant impact on detector sensitivities. In models based on the modal approach above, the laser beam interrogating the front face of a test mass is considered to sense the displacement resulting from the incoherent sum of the thermal displacements in the 'tails' of the test mass resonant modes (see for example [91]). However recent work by Levin [92], and others [93, 94] has highlighted the fact that models based on the modal approach above accurately describe the thermal noise in a test mass suspension only when the off resonance thermal noise from each mode is uncorrelated.

If the mechanical dissipation in a system is spatially inhomogeneous this assumption is no longer correct as correlations exist between the noise from the modes. Then the thermal displacement at any frequency can be greater or smaller than a modal treatment would suggest [92, 95, 96]. Using Levin's approach the power spectral density of thermally excited displacement of the front face of a test mass mirror may be written as

$$S_x(f) = \frac{2k_B T}{\pi^2 f^2} \frac{W_{diss}}{F_o^2}$$

where $W_{diss}$ is the power dissipated when a notional oscillatory force of peak magnitude $F_0$ acting on the face of a test mass mirror results in a pressure of spatial profile identical to that of the laser beam used to sense the mirror displacement. It can be shown that

$$W_{diss} = 2\pi f \int_{vol} \varepsilon(x,y,z) \phi(x,y,z,f) dV$$

where $\varepsilon$ is the energy density of elastic deformation when the test mass is maximally deformed under the applied notional pressure. Since, in this model, the power dissipated is a direct result of the elastic displacement caused by the notional pressure on the front face of the mirror, it can be seen that dissipation physically located close to the front face will contribute more to the dissipated power, and hence to the resulting thermal displacement, than dissipation located far from the front face.

This realisation is having considerable consequences for the design of test mass suspensions for the next generation of gravitational wave detectors and highlights the importance of a source of dissipation which has been the subject of much study in recent years – that of the ion-beam-sputtered multi-layer dielectric mirror coatings which must be applied to the front faces of the test masses to form highly reflecting mirrors. However, before considering mirror coatings, attention must first be turned to the choice of a suitable substrate material.

In the coated suspended test masses thermal noise appears in two forms, what we will call Brownian thermal noise resulting from the internal friction of the materials forming the test masses, coatings and suspension elements, and thermoelastic thermal noise. The latter results from thermodynamic fluctuations of temperature in the suspension systems, which then effectively couple to displacements predominantly through the expansion coefficient of the materials [97, 98]. Both Brownian and thermoelastic dissipation can be of a level significant for interferometric detectors.

All current interferometric detectors use fused silica as a mirror substrate material. This choice is a result of fused silica having suitable optical properties in addition to relatively low Brownian dissipation with many research groups having contributed to the study of the level and sources of dissipation in silica. Experimental measurements from Numata [99] suggested that an interesting frequency dependence in the loss factor of silica was evident, making the predicted thermal noise from silica substrates better at low frequencies than had previously been assumed. Work from a number of researchers suggests that a subset of Suprasil fused silica (grades 311 and 312) has consistently lower mechanical loss than other available silicas, with loss factors lower than $10^{-8}$ having been measured [100]. Penn has developed a semi-empirical model for the expected level of measured dissipation in fused silica samples with contributions to dissipation arising from internal frictional losses in both the bulk and surface layers of the samples [101]. This model predicts that for substrates of a geometry desirable for detectors such as Advanced LIGO expected dissipation may be substantially lower than previously predicted. Experiments are ongoing to verify these predictions.

An alternative material of current interest for transmissive substrates is sapphire. Sapphire has been demonstrated to have extremely low Brownian dissipation at frequencies above the gravitational wave detection band with dissipation factors as low as ~ 2 x $10^{-9}$ having been measured at room temperature [102]. However as recently pointed out by Braginsky and colleagues [98], the thermo-mechanical properties of sapphire are such that the thermoelastic thermal noise from sapphire test mass substrates can be significantly higher in the gravitational wave detection band than thermal noise from Brownian dissipation.

Sapphire currently forms the baseline choice for the substrate material for the planned Advanced LIGO detector upgrade. However trade-offs between the expected thermal noise performance of these substrate materials against other considerations such as the power handling capability, availability, etc of the substrates are a subject of current study.

As mentioned above dissipation associated with the dielectric mirror coatings added to the test masses is of particular significance for Advanced LIGO and future generations of interferometric detectors. From experiment and calculation, significant sources of coating mechanical dissipation exist in the form of Brownian dissipation intrinsic to the coating materials [103, 104, 105, 106, 107] and from a form of thermoelastic dissipation in the case where the thermo-mechanical properties of the coating multi-layers are different from those of the substrate. [108, 109]. In both cases the magnitude of the resulting thermal noise depends on the relative properties of the coating and substrate, hence for example, the optimal coating for one substrate material may not be optimal for another.

Studies of the most commonly used type of coating, formed from alternating multi-layers of ion-beam-sputtered $SiO_2$ and $Ta_2O_5$, suggest strongly that the coating dissipation is dominated by the $Ta_2O_5$ component of the coating [106, 107]. Current research is thus targeted at identifying ways to reduce the dissipation of $Ta_2O_5$ or finding an alternate high index material.

Reduction of coating thermal noise forms a significant challenge to be overcome in designing generations of interferometers to follow Advanced LIGO.

*4.3 Photoelectron shot noise and the quantum limit*

For gravitational wave signals to be detected, the output of the interferometer must be held at one of a number of possible points on an interference fringe. While an obvious point to choose is halfway up a fringe since the change in photon number produced by a given differential change in arm length is greatest at this point, it can be shown that the best signal-to-noise ratio is obtained as the locking point approaches the bottom of the fringe [110]. The interferometer may be stabilised to the required point on a fringe by sensing any changes in intensity at the interferometer output with a photodiode and feeding the resulting signal back, with suitable phase and dc bias, to a transducer capable of changing the position of one of the interferometer mirrors. Information about changes in the length of the interferometer arms can then be obtained by monitoring the signal fed back to the transducer.

As mentioned earlier it is very important that the system used for sensing the optical fringe movement on the output of the interferometer can resolve strains in space of 2 x$10^{-23}$/√Hz or lower, or differences in the lengths of the two arms of less than $10^{-19}$ m/√Hz, minute displacements compared to the wavelength of light ($10^{-6}$ m). A limitation to the sensitivity of the optical readout scheme is set by shot noise in the detected photocurrent. From consideration of the number of photoelectrons (assumed to obey Poisson statistics) measured in a time $t \sim 1/2\Delta f$ it can be shown [110] that the detectable strain sensitivity depends on the level of laser power, $P$, of wavelength $\lambda$ used to illuminate the interferometer of arm length $L$, and over a bandwidth $\Delta f$, such that:

$$\delta x^2 \cong \frac{\hbar c \lambda}{4\pi \, P \cos^2(\phi/2)} \Delta f$$

$$\cong \frac{\hbar c \lambda}{4\pi \, P} \Delta f \quad \text{when } \phi = 0$$

where $c$ is the velocity of light, $\phi$ is the phase difference between the light in the two arms of the interferometer, and $\hbar$ is Planck's constant. We assume that the photodetectors have a quantum efficiency ~1. It should be noted that the best sensitivity is not actually obtained by locking half way up a fringe but by operating close to the point where $\phi \sim 0$ or the output intensity is zero. Achievement of the required strain sensitivity level requires a laser, operating at a wavelength of $10^{-6}$ m, to provide 6 x $10^6$ W power at the input to a simple Michelson interferometer. This is a formidable requirement; however there are a number of techniques which allow a large reduction in this power and these were discussed earlier in terms of the need for delay lines or cavities in the arms and the advantages of power recycling.

Such power levels also produce fluctuations in radiation pressure on the mirrors and it can easily be shown that in the case of a simple Michelson interferometer the resulting equivalent differential displacement sensitivity is given by

$$\delta x^2 \cong \frac{\hbar P}{\lambda m^2 \pi^3 f^4 c} \Delta f$$

where $m$ is the mass of each end mirror of the interferometer. For ease of calculation we have assumed that the beamsplitter has infinite mass.

If the photon noise fluctuations are statistically independent of the radiation pressure fluctuations – a valid assumption in the case of the simple Michelson as they have been shown [111, 112] to arise from orthogonal fluctuations of the vacuum field entering the unused port of the beamsplitter – then the two effects can be combined additively to give

$$\delta x^2 \cong \frac{\hbar c \lambda}{4\pi \ P} \Delta f \ + \ \frac{\hbar P}{\lambda m^2 \pi^3 f^4 c} \Delta f$$

Clearly there is an optimum frequency dependent operating power to give minimum noise level and in this situation the minimum detectable noise spectral amplitude of displacement is

$$\delta x^2 \cong \frac{\hbar}{m \pi^2 f^2} \Delta f$$

This argument can be generalized for multiple beams in the arms and for Fabry-Perot cavities and essentially the same result is obtained. This is really an example of the Heisenberg Microscope experiment and thus it is not surprising that the same result can be obtained by using the Heisenberg Uncertainty Principle to calculate the uncertainty in position of the interferometer test masses (110). This apparent limitation to sensitivity is known as the Standard Quantum Limit (SQL).

It is important to note that the above calculation relies on the lack of correlation between the displacement limits set by the photon noise and those set by the radiation pressure noise. There are a number of interesting corollaries to this.

Firstly if it is possible to alter the balance of the fluctuations in the two quadratures of the vacuum field it is possible to reach the SQL at lower power levels than required in the above analysis. Such an imbalance can be achieved by 'squeezing' the vacuum fluctuations entering the unused port of the beamsplitter [112]. Squeezing has been experimentally demonstrated in a number of laboratories, see for example [113, 114, 115, 116], but of particular note are recent results from McClelland and colleagues in Australia [117] who have demonstrated several dBs of squeezing at the frequencies relevant for ground based gravitational wave detectors. Further, if correlations are present between the displacement limits discussed above, it is possible, at least in principle, to bypass the limit set by the SQL [118]. There are at least two ways to introduce such correlations:
- Through using a cavity configuration where there is a strong optical spring effect coupling the optical field to the mechanical system. Such effects can be enhanced by using intra-cavity readout schemes where the motion of small internal test masses is monitored with a local transducer which might use microwaves rather than light. Such schemes – optical bars, optical levers, and symphotic states – have been devised and studied in depth by Braginsky and colleagues at the University of Moscow (see for example [119, 120]).
- Through measuring the output signal after suitably designed filtering at optical frequencies – filtering, by means of long Fabry-Perot cavities, which effectively introduce correlations [121,122].

Of course another possibility to evade the SQL is to measure a different variable, one for which the measurement operator commutes with the operator resulting from the back action. This implies that the measurement operator at one time should commute with itself at a later time. Clearly '*x*' - displacement - is not such an operator as '*x*' at one time is correlated with '*x*' at a later time through the HUP relationship with momentum '*p*'. However '*p*' is a suitable operator as although a measurement of '*p*' results in an uncertainty in '*x*', this does not feed back into '*p*'. Thus if a velocity measurement system - speedmeter - is devised, this allows performance below the SQL. A number of systems have been suggested for speedmeters (see for example [121, 123,124,125]), the most straightforward being the implementation of a Sagnac configuration [126].

It should be noted that the signal-recycling concept as currently used in GEO 600 and planned for Advanced LIGO has the potential of allowing measurements below the SQL [127]. The asymmetry introduced by narrow-banding the sensitivity offset on one side of the optical carrier introduces a correlation between the photoelectron shot noise and the effect of the back-reaction. In this case quantum noise curves of the type included in figure 5 have been calculated. It is interesting to note that at its lowest point the quantum noise is better than would be predicted by the SQL. In principle the quantum noise limited sensitivity at different frequencies may be further improved by using squeezed light for illumination of the system and/or by using a long filtering cavity before the detection of the signal out of the system [128, 129].

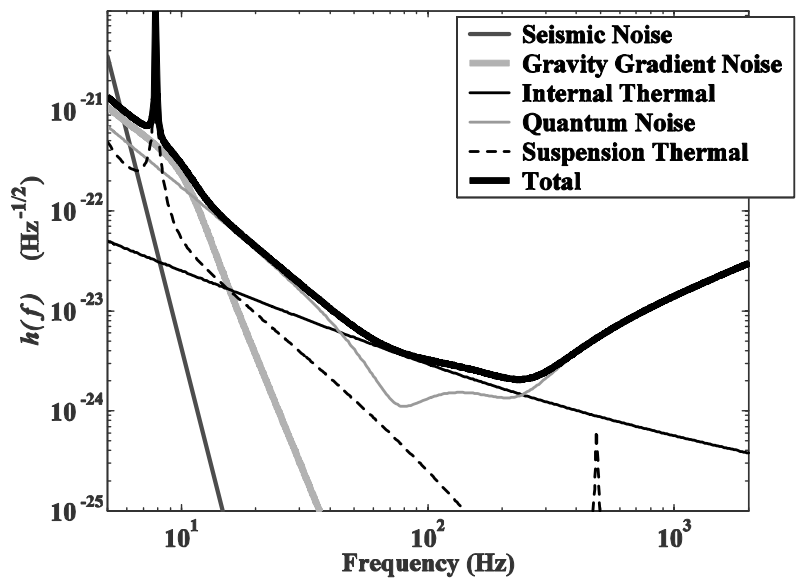

**Figure 5.** Sensitivity curve, showing noise anatomy, for the planned Advanced LIGO detector system.

These techniques for sensitivity enhancement beyond the SQL require losses in all parts of the main optical system to be very low and the quantum efficiencies of the photo-detection systems to be very high; otherwise the noise resulting from the statistical fluctuations associated with the loss tend to wash out the improvements obtained.

## 5. Future Detectors

The next stage forward in interferometric detectors is well defined with the design for Advanced LIGO incorporating silica fibre suspension, signal recycling and higher power lasers being well advanced. There are factors associated with coatings that make further development of the Advanced LIGO design difficult. These are the mechanical loss factors – and thus associated thermal/thermoelastic noise – of the dielectric mirror coatings (as discussed earlier), and also the optical absorption of the coatings as the heat then dumped in the test masses, particularly the input test masses of the cavities results in thermal effects which are difficult to compensate for.

There are a number of different ways forward. Materials developments may well improve both the mechanical loss and the optical absorption of the coatings. Further, cooling of the test masses may reduce the thermal noise problem due to the coatings and this approach is being adopted in Japan [130] for their new proposed long baseline detector, LCGT, which is intended to have the same performance as Advanced LIGO.

For future detectors non-transmissive optics are being mooted to bypass the problem of thermal distortion of light passing through an input cavity optic – see [131, 132, 133] and this matches well with proposals [134, 135] to use silicon as a test mass material motivated by its low mechanical loss and high thermal conductivity. Of course squeezed light techniques may be useful also in reducing the required laser power for a given sensitivity, thus helping with the distortion problem. Once the material problems have been alleviated the use of techniques to bypass the SQL immediately become attractive. Design tradeoffs between different techniques will be an area of fertile research for the next ten years.

Another very active field of detector research is that of preparing for a space borne experiment and the principles of LISA (Laser Interferometer Space Antenna) are outlined in the next section.

*5.1 Longer baseline detectors in space*

Some of the most interesting gravitational wave signals (those resulting from the formation and coalescence of black holes in the range $10^3$ to $10^6$ solar masses) will lie in the region of $10^{-4}$ Hz to $10^{-1}$ Hz. To search for these requires a detector whose strain sensitivity is approximately $10^{-23}$ over relevant timescales. Approximately 25 years ago Bender, with colleagues in Boulder [136], pointed out that the most promising way of looking for such signals is to fly a laser interferometer in space, i.e. to launch a number of drag free spacecraft into orbit and to compare the distances between test masses in these craft using laser interferometry. The JILA based project LAGOS (Laser Antenna for a Gravitational Observatory in Space) developed during the 1980s [137, 138] and then accelerated in the 1990s through increased European interest in proposing to ESA a space-based detector. LISA (Laser Interferometer Space Antenna) [6,7] was born and is now a joint ESA/NASA mission, being developed by a multinational research team. LISA consists of an array of three drag free spacecraft at the vertices of an equilateral triangle of length of side 5 x $10^6$ km. This cluster is placed in an Earth-like orbit at a distance of 1 AU from the Sun, and 20 degrees behind the Earth as shown in figure 6. Proof masses inside the spacecraft (two in each spacecraft) form the end points of three separate but not independent interferometers.

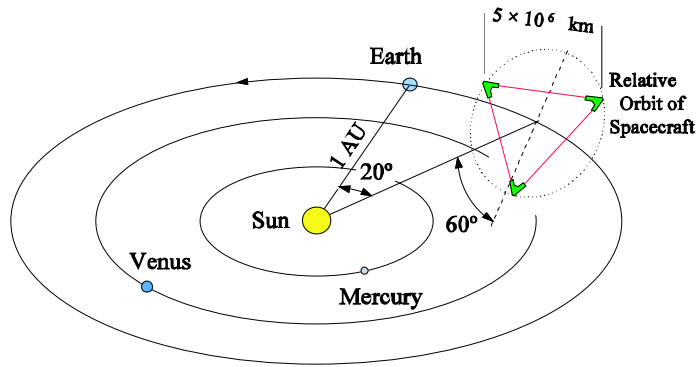

**Figure 6.** Schematic diagram of LISA and its orbit about the sun

The three-interferometer configuration provides redundancy against component failure, gives better detection probability, and allows determination of the polarisation of incoming radiation. The spacecraft accommodating the proof masses shields them from external disturbances. Drag free control servos enable the spacecraft to follow the proof masses to a high level of precision, the drag compensation being effected using proportional electric thrusters. Requirements on acceleration noise are very tight – a few times $10^{-15}$ ms$^{-2}$/√Hz at $10^{-4}$ Hz. This requirement is many orders of magnitude better than previously flown drag-free systems [139, 140] and has led to the requirement for a single satellite mission to demonstrate that performance within an order of magnitude of what is finally required can be achieved. This demonstrator mission – LISA Pathfinder will carry drag free packages from Europe (Lisa Test Package or LTP [141]) and from the USA (ST7 [142]).

The interferometry in LISA has to be different from that in the ground-based detectors because the arm lengths are not stationary and thus the light returning to a spacecraft is Doppler shifted from that which left the craft and thus heterodyne signals are produced. Operation of LISA requires that the phase differences of the incoming and outgoing light at each spacecraft are measured and the results combined to extract the relevant signals. Just as in the case of the ground based detectors, the presence of laser frequency noise is a limiting factor as is the frequency noise is the clocks used to beat down the heterodyne signals for phase measurement. Algorithms to minimize the effect of these noise sources which involve combining the phase measurements from the arms with different time delays – the process is known as Time Delay Interferometry (TDI) – is too detailed for this article and full information can be found in the literature [143, 144, 145].

LISA is expected to be launched around 2013 and to be producing data for up to ten years thereafter. Follow on missions to extend the frequency range and to allow different sources to be targeted are already being proposed. One such is 'Big Bang Observer' involving multiple constellations to allow cross correlations in the search for stochastic background signals.

## 6. Some Binary Systems and Prospects for Detection

In this section we shall discuss radiation from coalescing binary systems of particular relevance to both ground based and space based detectors and also massive black hole binaries, of particular relevance to space based systems. We refer the reader to recent reviews [5, 146, 147, 148] for further reading on other types of sources. A compact binary, consisting of neutron stars (NS) and/or black holes (BH), evolves by emitting gravitational radiation which extracts the rotational energy and angular momentum from the system, thereby leading to an inspiral of the two bodies towards each other. The dynamics of a compact binary consists of three phases: (i) The *early inspiral phase* in which the system spends 100's of millions of years and the power emitted in GW is low. The signal has a characteristic shape with slowly increasing amplitude and frequency and is called a *chirp* waveform. A binary signal that chirps (i.e. its frequency changes perceptibly during the course of observation) is an astronomer's *ideal standard candle* [149] and by observing the radiation from a *chirping* binary we can measure the luminosity distance to the source. (ii) The *merger phase* when the two stars are orbiting each other at a third of the speed of light and experiencing strong gravitational fields with the gravitational potential being $\phi = GM/Rc^2 \sim 0.1$. This phase warrants the full non-linear structure of Einstein's equations as the problem involves strong relativistic gravity, tidal deformation (in the case of BH-BH or BH-NS) and disruption (in the case of BH-NS and NS-NS) and has been the focus of numerical relativists [150] for more than two decades. (iii) The *late merger phase* when the two systems have merged to form either a single NS or a BH, settling down to a quiescent state by radiating the deformations inherited during the merger. The emitted radiation can be computed using perturbation theory and gives the quasi-normal modes (QNM) of BH and NS. The QNM carry a unique signature that depends only on the mass and spin angular momentum in the case of BH, but depends also on the equation-of-state (EOS) of the material in the case of NS. The three phases help to test general relativity in ways that will shed new insights into the non-linear structure of general relativity.

*6.1 NS-NS binaries*

Radio astronomers have observed three NS binaries in our own Galaxy that will coalesce within the Hubble time. Based on the observations of such binaries it has been estimated that Galactic coalescence rate of double NS is $\sim 9 \times 10^{-5}$ yr$^{-1}$. [151, 65]. NS binaries should be seeable in initial and Advanced LIGO detectors to 20 Mpc and 300 Mpc, respectively (cf. Figure 6), which would imply an event rate of NS-NS coalescences of up to 0.1 and 500 yr$^{-1}$, respectively. Although current state-of-the-art theoretical waveforms will serve as good templates for detection detailed relativistic hydrodynamical simulations (see, e.g. Ref. [152]) would be needed to interpret the emitted radiation during the coalescence phase, wherein the two stars collide to form a bar-like structure prior to merger. Over a couple of dynamical time-scales the bar deformity decays, emitting strong bursts of GW. Observing the radiation from this phase should help us to deduce the equation-of-state (EOS) of NS bulk matter.

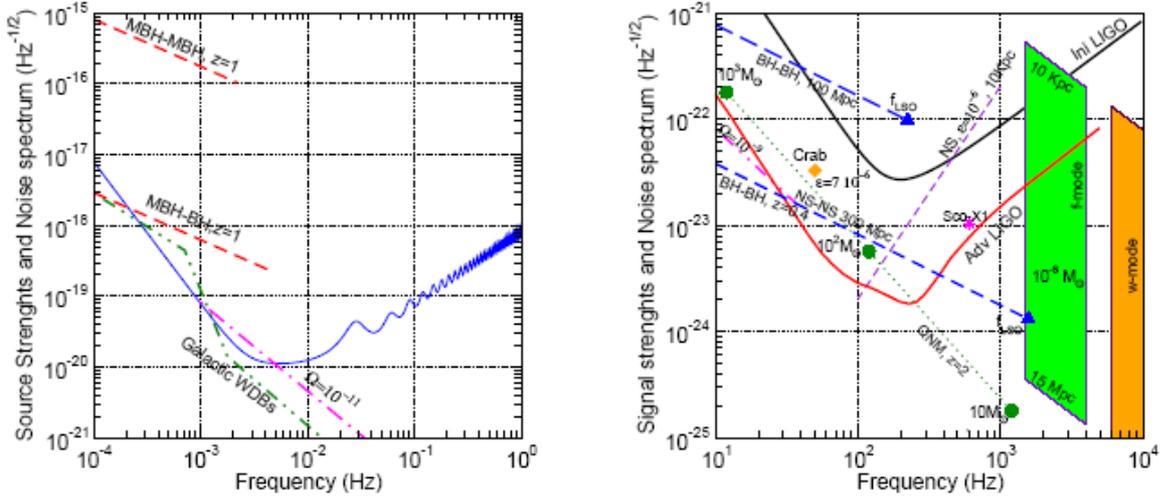

**Figure 7:** The amplitude spectrum of LISA (left panel) and Initial and Advanced LIGO (right panel) are shown together with the strengths of various sources. See text for details.

*6.2. NS-BH binaries*

Advanced interferometers will be sensitive to NS-BH binaries out to a distance of about 650 Mpc. The rate of coalescence of such systems is not known empirically as there have been no astrophysical NS-BH binary identifications. However, the population synthesis models give [147] a Galactic coalescence rate in the range $3 \times 10^{-7} - 5 \times 10^{-6}$ yr$^{-1}$. The event rate of NS-BH binaries will be worse than BH-BH of the same total mass by a factor of $(4\eta)^{3/2}$ since the SNR goes down as $\sqrt{4\eta}$. $\eta$ is the mass ratio of the two components of mass $m_1$ and $m_2$, and is defined as $\eta = m_1 m_2/(m_1+m_2)^2$ Taking these factors into account we get an optimistic detection rate of NS-BH of 1 to 1500 in initial and Advanced LIGO, respectively. NS-BH systems are very interesting from an astrophysical point of view as it might be possible to measure the radius of the NS to ~15% and thereby infer the EOS of NS [153].

*6.3. BH-BH binaries*

The span of initial LIGO to BH-BH binaries will be about 150 Mpc and of Advanced LIGO a red-shift of $z = 0.4$–0.55. As in the case of NS-BH binaries, here too there is no empirical estimate of the event rate. Population synthesis models are highly uncertain about the Galactic rate of BH-BH coalescences and predict [147] a range of $3 \times 10^{-8}$–$10^{-5}$ yr$^{-1}$, which is smaller than the predicted rate of NS-NS coalescences.
However, owing to their greater masses, BH-BH event rate in our detectors is larger than NS-NS by a factor $M^{5/2}$ for M $\leq$ 100M$_{sun}$. The predicted event rate is a maximum of 1 yr$^{-1}$ in initial LIGO and 500 yr$^{-1}$ to 20 day$^{-1}$ in Advanced LIGO. Black hole mergers are the most promising candidate sources for a first direct detection of GW. The two BHs experience the strongest possible gravitational fields before they merge with each other and serve as a platform to test general relativity in the non-linear regime. For instance, one can detect the scattering of GW by the curved geometry of the binary [154], and measure, or place upper limits on, the mass of the graviton to $2.5 \times 10^{-22}$ eV and $2.5 \times 10^{-26}$ eV in ground- and space-based detectors, respectively [155]. High SNR events (which could occur once every month in Advanced LIGO) can be used to test the full non-linear gravity by comparing numerical simulations with observations and thereby gain a better understanding of the two-body problem in general relativity.

*6.4 Massive black hole binaries*

It is now believed that the centre of every galaxy hosts a BH whose mass is in the range $10^6$–$10^9 M_{sun}$ [156]. These are termed as *massive black holes* (MBH). There is now observational evidence that when galaxies collide the MBH at their nuclei might get close enough to be driven by gravitational radiation reaction and merge within the Hubble time [157]. Supermassive BH mergers will appear as the most spectacular events in LISA requiring no templates for signal identification, although good models will be needed to extract source parameters. Mergers can be seen to z ~ 30 and, therefore, one could study the merger-history of galaxies throughout the Universe and address astrophysical questions about the origin, growth and population of MBH. The recent discovery of a MBH binary [19] and the association of X-shaped radio lobes with the merger of MBH [158] predicts a rate for MBH mergers to be about 1 $yr^{-1}$ out to a red-shift of z = 5 [159].

*6.5 Equation-of-State and normal modes of neutron stars*

To determine the equation of state (EOS) of a neutron star, and hence its internal structure, it is necessary to independently determine its mass and radius. Astronomical observations cannot measure the radius of a neutron star, although radio and X-ray observations do place a bound on its mass. Therefore, it has not been possible to infer the EOS. Neutron stars will have their own distinct normal modes and GW observations of these modes should resolve the matter here since by measuring the frequency and damping times of the modes it would be possible to infer both the radius and mass of NS. The technique is not unlike helioseismology where observation of normal modes of the Sun has facilitated insights into its internal structure. In other words, GW observations of the normal modes of the NS will allow gravitational asteroseismology. Figure 7 shows in two shaded regions the amplitude of expected radiation from f- and w-modes, assuming a dissipation of $10^{-6} M_{sun}$ into the modes, for a NS located at within our galaxy (highest amplitudes) and at the Virgo supercluster (lowest amplitudes). The depicted range of frequency corresponds to different EOS and GW observations should essentially measure the EOS.

**7. Conclusion**

Some early relativists were sceptical about the existence of gravitational waves; however, the 1993 Nobel Prize in Physics was awarded to Hulse and Taylor for their experimental observations and subsequent interpretations of the evolution of the orbit of the binary pulsar PSR 1913+16, the decay of the binary orbit being consistent with angular momentum and energy being carried away from this system by gravitational waves. Thus it is now universally accepted that gravitational waves must exist unless there is something seriously wrong with General Relativity. There are many significant experiments underway and at the planning stage and the community is poised now to herald their detection and the start of a new astronomy. For the reader who wishes further information, we recommend a number of reviews/books spanning the last thirty years [160-164].

**8. Acknowledgements**

We wish to thank our colleagues in all the gravitational wave experiments for useful discussions and for the use of some of their figures. We are particularly grateful to Peter Fritschel at MIT for the Advanced LIGO noise anatomy curves. We are indebted to PPARC, the University of Glasgow and Cardiff University for support.

**List of figure captions**

**Figure 1.** Schematic diagram of how gravitational waves interact with a ring of matter. The 'quadrupole' nature of the interaction can be clearly seen, and if the mirrors of the Michelson Interferometer on the right lie on the ring with the beamsplitter in the middle, the relative lengths of the two arms will change and thus there will be a changing interference pattern at the output.

**Figure 2.** Predicted signal strengths for a number of possible sources of gravitational waves

**Figure 3.** (a) Room temperature bar detector in Munich (with H. Billing) and (b) Low temperature bar detector, Auriga, at Legnaro.

**Figure 4.** Schematic diagram and birds eye view of LIGO (Hanford).
Image courtesy of the LIGO Science Collaboration.

**Figure 5.** Sensitivity curve, showing noise anatomy, for the planned Advanced LIGO detector system.

**Figure 6.** Schematic diagram of LISA and its orbit about the sun.

**Figure 7:** The amplitude spectrum of LISA (left panel) and Initial and Advanced
LIGO (right panel) are shown together with the strengths of various sources. See text
for details.